\documentclass[11pt]{article}
\usepackage{amsmath}
\usepackage{amsgen,amstext,amsbsy,amsopn,amsthm, amssymb}
\usepackage{amsmath,amsfonts,amsthm,amssymb}

\newcommand\version{December 17, 2007}
\numberwithin{equation}{section}

\newcommand\x{x}
\newcommand\y{y}

\newcommand{\beq}{\begin{equation}}
\newcommand{\eeq}{\end{equation}}

\newcommand{\beqa}{\begin{eqnarray}}
\newcommand{\eeqa}{\end{eqnarray}}
\newcommand\Z{{\mathbb Z}}

\newcommand\N{{\mathbb N}}

\newcommand\R{{\mathbb R}}
\newcommand\eps{\varepsilon}
\newcommand\half{\mbox{$\frac 12$}}

\newcommand\rhov\varrho

\newcommand\const{{\rm const.\, }}
\newcommand\Tr{{\rm Tr}}

\newcommand\Hh{{\cal H}}

\newcommand\infspec{{\rm inf\, spec\, }}

\renewcommand\aleph\varaleph
\renewcommand\rho\varrho

\def\Co#1{C_{#1}^{\phantom{\dagger}}}
\def\Cod#1{C_{#1}^\dagger}
\newcommand\dt{\Xi}
\newcommand\rhoup{\rho_u}
\newcommand\rhodown{\rho_d}

\newtheorem{thm}{THEOREM}

\newtheorem{lem}{LEMMA}
\newtheorem{cor}{COROLLARY}

\begin{document}

\markboth{\scriptsize{\version}}{\scriptsize{\version}}
\title{\bf{Ground state energy of the low density Hubbard model}}
\author{\vspace{5pt} Robert Seiringer and  Jun Yin
\\ \vspace{-2pt}\small{ Department of Physics, Jadwin Hall, Princeton
University,  }\\ \vspace{-2pt}\small Princeton, NJ 08542-0708, USA
  \\ {\small Email: \texttt {\{rseiring/junyin\}@princeton.edu}} }

\date{\small \version}
\maketitle

\begin{abstract}
  We derive a lower bound on the ground state energy of the Hubbard
  model for given value of the total spin. In combination with the
  upper bound derived previously by Giuliani \cite{A}, our result
  proves that in the low density limit, the leading order correction
  compared to the ground state energy of a non-interacting lattice Fermi gas is
  given by $8\pi a \rho_u \rho_d$, where $\rho_{u(d)}$ denotes the
  density of the spin-up (down) particles, and $a$ is the scattering
  length of the contact interaction potential. This result extends
  previous work on the corresponding continuum model to the lattice
  case.
 \end{abstract}

\renewcommand{\thefootnote}{${\, }$}
\footnotetext{\copyright\,2007 by the authors.
This paper may be reproduced, in its entirety, for non-commercial
purposes.}

\section{Introduction}

In recent years, much effort has been made to rigorously analyze the
properties of dilute quantum gases at low temperature and low
density. For repulsive pair interaction potentials, the leading order
correction compared to the case of ideal quantum gases of the ground
state energy and free energy of continuous quantum gases in the
thermodynamic limit have been investigated in
\cite{dyson,LY1,LY2,LSY00,LSS,LSSY,S06,S2006}.  

In particular, in \cite{LSS} it was proved that the ground state
energy per unit volume of a low-density spin $1/2$ Fermi gas with repulsive pair interaction is
(in units where $\hbar=2m=1$) given by
\begin{equation}\label{lssr}
\frac 35 \left( 6\pi^2 \right)^{2/3}
\left(\rho_u^{5/3}+\rho_d^{5/3}\right) + 8\pi a \rho_u \rho_d +
o(\rho^2)\,.
\end{equation}
Here, $\rho_{u(d)}$ denotes the density of the spin-up (down)
particles, and $a>0$ is the scattering length of the interaction
potential. The total density of the system equals $\rho=\rho_u +
\rho_d$, and $(\rho_u-\rho_d)/(2\rho)$ is the average spin per
particle.

The goal of this paper is to extend the analysis in \cite{LSS} from
the continuum to the lattice case. We restrict our attention to the
case of a simple cubic, three-dimensional lattice. Without loss of
generality, we choose units such that the spacing between two
neighboring lattice sites is one; i.e., the configuration
space for one (spinless) particle is $\Z^3$.

For simplicity, we consider the case where the interaction potential
between the particles has zero range, i.e., only particles on the same
lattice site interact. This is the simplest version of the {\it
  Hubbard model}, which was originally introduced as a highly
simplified model for fermions with repulsive (Coulomb)
interaction. For a review of the history, rigorous results and open
problems, we refer to \cite{T,L}.

An upper bound to the ground state energy of the desired form was
already derived in \cite{A}, hence we concentrate here on the lower
bound. Our main result is given in Theorem~\ref{thm1} below. In
combination, the two bounds show that the ground state energy per unit volume of the
Hubbard model at low density $\rho=\rho_u+\rho_d$ and at given spin
polarization $(\rho_u-\rho_d)/(2\rho)$ is given by
$$
e_0(\rho_u, \rho_d)+8\pi a\rho_u\rho_d+o(\rho^2)\,,
$$
where, as before, $a$ denotes the (appropriately defined) scattering
length of the interaction potential, and $e_0(\rho_u,\rho_d)$ is the
ground state energy per unit volume of the ideal lattice Fermi gas.

\subsection{Model and Main Result}

We consider particles hopping on the simple cubic lattice $\Z^3$. For
$N$ spinless fermions, the appropriate Hilbert space $\Hh(N)$ is the
subspace of totally antisymmetric functions in $L^2(\Z^{3N})$, with
norm 
$$
\|\psi\|_2=\sqrt{\sum_{x_1\in \Z^3}\cdots\sum_{x_N\in
    \Z^3}|\psi(x_1,\dots x_N)|^2}\,.
$$
We define $\mathcal H(N_u, N_d)\subset L^2(\Z^{3N_u+3N_d})$ as
\begin{equation}\label{Hud} 
\mathcal H(N_u, N_d) = \mathcal H(N_u)\otimes
\mathcal H(N_d) \,.
\end{equation}
Elements of $\Hh(N_u, N_d)$ are thus functions of $N_u+N_d$ variables
that are antisymmetric both in the first $N_u$ and the last $N_d$
variables.  In appropriate units, the Hubbard Hamiltonian for
$N_u$ spin-up particles and $N_d$ spin-down fermions in a box
$[0,L]^3$ is given by 
\begin{equation}\label{defH}
H =-\sum_{i=1}^{N_u}\Delta_{x_i}-\sum_{i=1}^{N_d}\Delta_{y_i}+ g \sum_{i=1}^{N_u}\sum_{j=1}^{N_d}
\delta_{x_i,y_j}\,, 
\end{equation}
where $x_1,\dots x_{N_u}$ and $y_1,\dots y_{N_d}$ are the coordinates
of the spin-up and spin-down particles, respectively. The usual
lattice Laplacian is denoted by $\Delta$; it acts as $(\Delta
f)(x)=\sum_{y,|x-y|=1}(f(y)-f(x))$.  We consider {\it Dirichlet
  boundary conditions} on $[0,L]$, i.e., we restrict $H$ to functions
that vanish outside the cube $[1,L-1]^3$. The thermodynamic limit
corresponds to taking $L\to \infty$, $N_u\to \infty$ and $N_d \to
\infty$ in such a way that
$$ 
\rho_u = \lim \frac{ N_u}{L^3} \quad {\rm and }\quad \rho_d = \lim \frac {N_d}{L^3}\,.
$$
Note that necessarily $0\leq \rho_{u(d)}\leq 1$ because of the
antisymmetry of the wavefunctions.

We remark that instead of considering two species of spinless fermions
with particle numbers $N_u$ and $N_d$, as we do here, one can
equivalently consider just one species of $N_u+N_d$ fermions with spin
$1/2$, and restrict to the subspace of total spin $S= (N_u-N_d)/2$. We
will use the former formulation for convenience.

The coupling constant $g$ is assumed to be nonnegative and is allowed
to take the value $+\infty$. The scattering length $a$ of the interaction
potential $g\delta_{0,x}$ is given by (c.f. \cite[Eq.~(1.6)]{A})
\begin{equation}\label{scl}
a=\frac{g}{8\pi(g\gamma+1)}\,, \quad 
\gamma=\int_{[-\pi,\pi]^3}\frac{1}{4\sum_{i=1}^3(1-\cos k^i)}\frac{dk}{(2\pi)^3} \,,
\end{equation}
where $k=(k^1,k^2,k^3)\in \R^3$. 
We refer to Section~\ref{sec:scl} for details.

Our main result of this paper is the following.

\begin{thm}\label{thm1}
  Let $E_0(N_u,N_d,L)$ be the ground state energy of $H$ in
  (\ref{defH}). If $L\to \infty$, $N_u\to \infty$, $N_d\to \infty$
  with $\rho_u=\lim_{L\to\infty}N_u/ L^3$ and
  $\rho_d=\lim_{L\to\infty}N_d/ L^3$, then 
\begin{equation}\label{mainresult}
  \liminf_{L\to\infty}\frac{1}{L^3}E_0(N_u,N_d,L)\geq e_0(\rho_u,
  \rho_d)+8\pi a\rho_u\rho_d\big( 1 - o(1)\big)\,,
\end{equation}
with $o(1) \leq C (a\rho^{1/3})^{1/15}$ for some constant $C>0$. 
 Here $e_0(\rho_u,
  \rho_d)$ is the ground state energy per unit volume of
  the ideal lattice Fermi gas, and $\rho=\rho_u+\rho_d$.
\end{thm}

As already pointed out in the Introduction, an upper bound of the
desired form (\ref{mainresult}) was proved by Giuliani in \cite{A}, extending the
method used in \cite{LSS}. This shows that Eq. (\ref{mainresult}) actually holds
as an equality.

It is easy to see that 
\begin{equation}\label{e0p} 
  e_0(\rho_u, \rho_d)=\frac 35 \left( 6\pi^2 \right)^{2/3} (\rho_u^{5/3}+\rho_d^{5/3})+ O(\rho^{7/3})
\end{equation}
for small $\rho$. Hence, at fixed $a$, the expression  for the
ground state energy of the continuous Fermi gas (\ref{lssr}) and the one
for the lattice Fermi gas coincide up to terms lower of order
$o(\rho^2)$. Theorem~\ref{thm1} is slightly stronger, however, since
the error term is $o(a\rho^2)$, not $o(\rho^2)$. Hence
(\ref{mainresult}) should be viewed as a result for small
$a\rho^{1/3}$, which could be achieved either by making $\rho$ small
or by making $a$ small. We point out, however, that the case of small
$a$ is much simpler than the case of small $\rho$, since the correct
result can be obtained via first order perturbation theory in this
case, while this is not possible for fixed $a$ and small $\rho$. For
small $a$ and fixed $\rho$, the Hartree-Fock approximation becomes
exact, as was shown in \cite{bach}.

We state and prove  Theorem~\ref{thm1} for the simple cubic
lattice and zero-range interaction, for simplicity, but the method can
be used in more general cases. For instance, different lattice
structures could be considered, or longer-ranged hopping. Interactions
of longer range could also be included, as long as they are
non-negative and have a finite scattering length; and particles with
more than two spin states could be considered, as in \cite{LSS}. In
combination with the methods developed in \cite{S06}, our technique
can also be applied at non-zero temperature to estimate the pressure
or the free energy.

The proof of Theorem~\ref{thm1} follows closely the corresponding
continuum result in \cite{LSS}, with several important and non-trivial
modifications, however. One of the main ingredients is the generalized
Dyson Lemma, stated in Lemma~\ref{lemmamain} below, which allows for
the replacement of the hard interaction potential $g\delta_{xy}$ by a
softer and longer ranged potential, at the expense of the high
momentum part of the kinetic energy. The proof of the corresponding
Lemma~4 in \cite{LSS} uses rotational invariance of $\R^3$ in an
essential way, and hence does not extend to the lattice case of
$\Z^3$, where such a symmetry is absent. (See also the discussion in
\cite{A}). Our new Lemma~\ref{lemmamain} does not rely on this
symmetry, however.
 
Another important estimate in \cite{LSS} that does not carry over to
the lattice case is a bound on the average number of particles
that are close to their nearest neighbor. The estimate in
\cite[Lemma~6]{LSS} uses an inequality by Lieb and Yau \cite[Thm.~5]{LYau}
whose proof also relies on the rotational invariance of $\R^3$. In
Lemma~\ref{lemlt} below we will present a weaker version of this inequality
which is equally valid in the lattice case.

The strategy of the proof of Theorem~\ref{thm1} is similar to the
proof in \cite{LSS}. First, we separate the Hamiltonian $H$ into two
parts, the low momentum part of the kinetic energy on the one hand,
and the high momentum part of the kinetic energy together with the
interaction energy on the other hand. The first part is larger than
$e_0(\rho_u,\rho_d)$, while the second part can be bounded from below
by a softer interaction potential; this is the content of
Lemma~\ref{lemmamain}. In Lemma~\ref{lemP}, we shall show that at low
density the one particle density matrix of the ground state of the
Hubbard model is close to a projection, namely the projection onto the
Fermi sea. This information will then be used to prove that the
expectation value of the softer interaction potential is given by
$8\pi a\rho_u\rho_d$ to leading order. To bound some of the error
terms, we need a bound on the expected number of particles whose
distance to the nearest neighbor is small. This will be accomplished
in Lemma~\ref{lemir}.

In the next section, we shall state some preliminaries and introduce
the notation used throughout the proof. Section~\ref{sec:lemmas} contains the
main three Lemmas, which we have already referred to above. Finally,
in Section~\ref{sec:proof} the proof of Theorem~\ref{thm1} will be given.

\section{Preliminaries}

\subsection{Notation}

We start by introducing some notation that will be used throughout the
proof. First, the gradient operator
$\nabla=(\nabla^1,\nabla^2,\nabla^3)$ on $L^2(\Z^3)$ is defined as
usual as
$$
(\nabla^i f) (x) = f(x+e^i) - f(x)\,,
$$
with $e^i$ denoting the unit vector in the $i$'th coordinate
direction. Its adjoint is given by $(\nabla^{i\dagger} f)(x) =
f(x-e^i) - f(x)$. The Laplacian can then be expressed in terms of the
gradient as
$$
- \Delta = \nabla^\dagger \cdot \nabla = \nabla \cdot \nabla^\dagger\,.
$$

For any subset $A\subset \Z^3$, we denote by $\theta_A$ its
characteristic function. It will be convenient to introduce the
notation
$$
\left[\nabla^\dagger\theta_A\nabla\right]_s \equiv
\half\left(\nabla^\dagger\cdot \theta_A\nabla+\nabla\cdot
  \theta_A\nabla^\dagger\right) \,.
$$
Note that this is a nonnegative operator which plays the role of the
(Neumann) Laplacian on $A$.  For $A\subset \Z^3$ bounded, we denote by
$P_A$ the projection onto the normalized constant function on $A$,
$$
P_A\equiv\frac{|\theta_A\rangle\langle \theta_A|}{\|\theta_A\|_2^2}\,.
$$

For $h\in L^1(\Z^3)$, we define the convolution operator $\Co{h}$ as 
$$
(\Co{h} \psi)(x)  = h*\psi(x) = \sum_{y\in \Z^3} h(x-y) \psi(y)\,.
$$
Its adjoint is given by $(\Cod{h} \psi)(x)  = \sum_{y\in \Z^3} \overline {h(y-x)} \psi(y)$. 

We recall also the natural definition of the Fourier transform,
mapping $L^2(\Z^3)$ to $L^2([-\pi,\pi]^3)$. For $p=(p^1,p^2,p^3) \in
\R^3$, $|p^i|\leq \pi$,
$$
\widehat{\psi}(p) = \sum_{x\in \Z^3}e^{-ip\cdot x}\psi(x) \,.
$$
Its inverse is given by
$$
\psi(x)=\frac{1}{(2\pi)^{3}}\int_{[-\pi,\pi]^3}e^{ip\cdot x}\widehat{\psi}(p)\, d p\,.
$$

Using the above definitions, the following properties are easily
verified: 
\begin{align}\label{kin} 
  \widehat{\delta_{x,0}}(p)&=1\\\nonumber
  \widehat{h*\psi}(p)&=\widehat h(p)\widehat\psi(p)\\\nonumber
  \|\psi\|_2^2&=\int_{[-\pi,\pi]^3}|\widehat{\psi}(p)|^2
  \frac{dp}{(2\pi)^{3}}\\\nonumber
  -\langle\psi|\Delta|\psi\rangle&=\sum_{j=1}^3\int_{[-\pi,\pi]^3}
  \left(2-2\cos p^{j}\right)|\widehat{\psi}(p)|^2\frac{dp}{(2\pi)^{3}} \,.
\end{align}
In particular, if $M$ and $M'$ are two functions satisfying
$$
|\widehat{M}(p)|^2+|\widehat{M'}(p)|^2=1 \quad \text{for all } p \in [-\pi,\pi]^3\,,
$$
we can decompose the Laplacian $\Delta$ as 
\begin{equation}\label{momsep} 
\Delta=
\Cod{M}\Delta \Co{M} + \Cod{M'} \Delta \Co{M'}\,.  
\end{equation}
We will use this decomposition in the proof of Theorem~\ref{thm1} in
order to separate the kinetic energy into the high momentum and the
low momentum parts.

Finally, it will be convenient to introduce the operator
\begin{equation}\label{defxi}
\dt_A\equiv[\nabla^\dagger\cdot\theta_A\nabla]_s+\theta_A\Delta\,.
\end{equation}
It has the property that 
\begin{equation}\label{equP1}
\langle f|\dt_A |g\rangle
=
\half\sum_{\substack{x\in A,y\notin A \\ |x-y|=1}}\left[f(x)+f(y)\right]\left[g(y)-g(x)\right]\,.
\end{equation}

\subsection{Scattering Length}\label{sec:scl}

We denote by $\varphi$ the solution of the zero-energy scattering equation 
\begin{equation}\label{sq}
-\Delta \varphi(\x)+\half g\delta_{0,\x}\varphi(\x)=0
\end{equation}
with boundary condition $\lim_{|\x|\to\infty}\varphi(\x)=1$. It is given by \cite[Eq.~(1.5)]{A}
\begin{equation}\label{scphi}
  \varphi(x) = 1 - 4\pi a \int_{[-\pi,\pi]^3} \frac{ e^{i p\cdot x}}{2 \sum_{j=1}^3 (1-\cos p^j)} \frac {dp}{(2\pi)^3}\,,
\end{equation}
where $a$ is the scattering length (\ref{scl}). It can be shown
\cite{martinsson} that there is a constant $C>0$ such that
\begin{equation}\label{decpp}
\left| \varphi(x) - 1 +\frac a{|x|} \right| \leq C \frac a{|x|^3}\,.
\end{equation}

Note that, in particular, $a =
\lim_{|\x|\to\infty}(\varphi(\x)-1)|\x|$.  It can be readily checked
that
\begin{equation}\label{ap}
-\sum_{x\in\Z^3}\Delta \varphi(\x)=\half g\varphi(0)=4\pi a \,.
\end{equation}
Another property of $\varphi$ we will need is \cite[Eq.~(1.7)]{A}
\begin{equation}\label{ap2}
\sum_{\substack{x\in A,y\notin A \\ |x-y|=1}} \left(\varphi(y)-\varphi(x)\right) = 4\pi a
\end{equation}
for any  simply connected domain $A$ containing the origin. 

\subsection {Non-interacting Fermions}

We recall here briefly the ground state energy of non-interacting
(spinless) fermions on the lattice $\Z^3$. For $p\in [-\pi,\pi]^3$,
the dispersion relation will be denoted by
$$
E(p) = 2 \sum_{i=1}^3 \left(1- \cos{p^i}\right)\,.
$$
For given density $0\leq \rho\leq 1$, the Fermi energy $E_{\rm f}(\rho)$ is determined by
\begin{equation}\label{deffe}
(2\pi)^{-3} \int_{E(p) \leq E_{\rm f}(\rho)} dp = \rho \,.
\end{equation}
The ground state energy per unit volume in the thermodynamic limit is then
$$
e(\rho) = (2\pi)^{-3} \int_{E(p) \leq E_{\rm f}(\rho)} E(p) \, dp\,.
$$
For spin $1/2$ particles with spin-up density $\rhoup$ and spin-down
density $\rhodown$, the ground state energy is thus
$e_0(\rhoup,\rhodown)=e(\rhoup)+e(\rhodown)$.

\section{Auxiliary Lemmas}\label{sec:lemmas}

\subsection {Lemma One}

As mentioned in the Introduction, Lemma~\ref{lemmamain} is the
main tools of this paper. It is similar to Lemma 4 in \cite{LSS}.  This
lemma allows for bounding the hard interaction $g\delta_{0,x}$ from below by
a softer interaction at the expense of the high momentum part of the
kinetic energy and some error terms.

\begin{lem}\label{lemmamain}
  For $r\in \N$, let $A(r)$ denote the cube $A(r)=[-r,r]^3\cap
  \Z^3$. For any function $h\in L^1(\Z^3)$ satisfying $1\geq
  \widehat h(p)\geq 0$, let
\begin{equation}\label{deff}
f_r(x)=\max_{y\in x+ A(r)}|h'(y)-h'(x)| \quad , \quad {\rm where\ } \widehat {h'}(p) = 1-\widehat h(p)\,.
\end{equation}
For $R\in \N$, let $U$, $W$ and $V$ denote the nonnegative operators 
\begin{equation}\label{defU}
U =  (2R+1)^{-3}\theta_{A(R)} \,,
\end{equation}
\begin{equation}\label{defW}
W =  16 \pi f_R\sum_{x\in \Z^3} f_R(x) 
\end{equation}
and
\begin{equation}
V = (2R+1)^{-3}[\theta_{A(R)}-P_{A(R)}]\,.
\end{equation}
There exists a constant $C>0$ such that for any $R\geq C$ and
$0<\eps<1$, $0<\eta<1$, 
\begin{equation}\label{mainlemma}
\Cod{h} [\nabla^\dagger \cdot \theta_{A(R)}\nabla]_s
\Co{h} +\frac{g}{2}\delta_{x,0}\geq4\pi
a\left[(1-\eps)(1-\eta)U-\frac{W}{\eps}- \frac{CV}{\eta}\right] \,.
\end{equation}
\end{lem}

Compared with the result in \cite[Lemma~4]{LSS}, there is an
additional error term $V$ on the right side of
(\ref{mainlemma}). There is no restriction that $U$ has to vanish at
the origin, however, which was necessary in \cite{LSS}.  We note that
the norm of $U$ is given by $\|U\|=(2R+1)^{-3}$, which is much smaller
than the norm of $(g/2)\delta_{x,0}$ (which is $g/2$) for our choice
of $R\gg 1$ below.

\begin{proof} 
We are actually going to prove the stronger statement that
\begin{equation}\label{tem2} 
\left\langle \Cod{h}
    [\nabla^\dagger\cdot\Theta\nabla]_s \Co{h}
    +\frac{g}{2}\delta_{x,0}\right\rangle_\psi\geq 4\pi
  a\left\langle(1-\eps)(1-\eta)U-\frac{W}{\eps}-
    \frac{CV}{\eta}\right\rangle_{\psi} 
\end{equation}
for any $\psi\in L^2(\Z^3)$. Here and in the following, we use the
shorthand notation $\langle \, \cdot \rangle_\psi = \langle\psi|\,
\cdot \, |\psi\rangle$.  The non-negative function $\Theta$ is defined
by
$$
\Theta = \frac 1{R-\widetilde R} \sum_{r=\widetilde R}^{R-1}\theta_{A(r)} \,,
$$
where we denote by $\widetilde R$ the largest integer less than $R/2$.
Since $\Theta \leq\theta_{A(R)}$, (\ref{tem2}) implies (\ref{mainlemma}).

To prove (\ref{tem2}), we first define $B_R$ as
$$
B_R = \langle\psi|
\Cod{h}[\nabla^\dagger\cdot\Theta\nabla]_s|\varphi\rangle+ \langle
\psi |\frac{g}{2}\delta_{x,0}|\varphi\rangle\,.
$$
Here, $\varphi$ is given in (\ref{scphi}). Using Schwarz's inequality,
if follows that $|B_R|^2$ is bounded from above as
\begin{equation}\label{tem3}
  |B_R|^2
  \leq \left \langle \Cod{h} [\nabla^\dagger\cdot\Theta\nabla]_s \Co{h} +\frac{g}{2}\delta_{x,0}\right\rangle_\psi \, \left\langle[\nabla^\dagger\cdot\Theta\nabla]_s+\frac{g}{2}\delta_{x,0}\right\rangle_\varphi \,.
\end{equation}
By the definition of $\varphi$,
$\left\langle[\nabla^\dagger\cdot\Theta\nabla]_s+\frac{g}{2}\delta_{x,0}\right\rangle_\varphi\leq
\left\langle -\Delta+\frac{g}{2}\delta_{x,0}\right\rangle_\varphi=
4\pi a$.  Hence we see that the left side of (\ref{tem2}) can be
bounded from below as
\begin{equation}\label{inequB}
\left\langle \Cod{h}[\nabla^\dagger\cdot\Theta\nabla]_s \Co{h}+\frac{g}{2}\delta_{x,0}\right\rangle_\psi \geq \frac{|B_R|^2}{4 \pi a}\,.
\end{equation}

Define $\chi\in L^2(\Z^3)$ via 
$$
|\chi\rangle =[\nabla^\dagger\cdot\Theta\nabla]_s
|\varphi\rangle+\Delta|\varphi\rangle=[\nabla^\dagger\cdot\Theta\nabla]_s
|\varphi\rangle+\frac{g}{2}\delta_{x,0}|\varphi\rangle\,.
$$
Alternatively, using (\ref{defxi}), $|\chi\rangle= (R-\widetilde
R)^{-1}\sum_{r=\widetilde R}^{R-1}\dt_{A(r)}|\varphi\rangle$. Hence
$\chi$ is supported in $A(R)\setminus A(\widetilde R-2)$. Moreover,
$\chi(x)$ is a non-negative function for $R$ large enough, as can be
seen from (\ref{equP1}) and the asymptotic behavior of $\varphi$ in (\ref{decpp}).

Let also
$|\alpha\rangle = [\nabla^\dagger\cdot\Theta\nabla]_s|\varphi\rangle$.
Since $\alpha(x) = \chi(x) - (g/2)\delta_{x,0} \varphi(0)$, also
$\alpha$ is supported on $A(R)$. Moreover, $\sum_x \alpha(x) = 0$
because of (\ref{ap}) and (\ref{ap2}).
Since $\Co{h}+ \Co{h'} = 1$, 
$$
B_R = \langle \psi|\chi\rangle - \langle\psi| \Cod{h'} |\alpha\rangle\,.
$$
Recall that $\sum_x \alpha(x) = 0$, and $\alpha$ is supported on $A(R)$. Hence
$$
\left|(\Cod{h'} \alpha)(x)\right| = \left| \sum_y \left( \overline{
      h'(y-x)} - \overline{h'(x)}\right) \alpha(y)\right| \leq f_R(x)
\sum_y |\alpha(y)| = 8\pi a f_R(x)\,,
$$
with $f_R$ defined in (\ref{deff}). Here we used the fact that
$\chi(x)$ is non-negative and supported away from the origin, hence
$\sum_x |\alpha(x)| = \sum_x \chi(x) + g \varphi(0)/2 = g \varphi(0) =
8\pi a$.

In particular, we conclude that 
$$
|B_R| \geq\left|\langle\psi|\chi\rangle\right|- 8\pi a \sum_{x\in \Z^3}|\psi(x)|f_R(x) \,.
$$
Using Schwarz's inequality and the definition of $W$ in (\ref{defW}),
we thus obtain
\begin{equation}\label{xxc}
|B_R|^2\geq (1-\eps)\langle\psi|\chi\rangle\langle\chi|\psi\rangle -
\frac{4\pi a^2}{\eps}\sum_{x\in \Z^3}|\psi(x)|^2 W(x)\,.
\end{equation}
To get a lower bound on $|\chi\rangle\langle\chi|$, we use again
Schwarz's inequality, as well as the fact that $\chi$ is supported on
$A(R)$. We obtain, for $0<\eta<1$,
\begin{align}\nonumber
  |\chi\rangle\langle\chi| &\geq
  (1-\eta)P_{A(R)}|\chi\rangle\langle\chi|P_{A(R)}
  +\left(1-\eta^{-1}\right)(\theta_{A(R)}-P_{A(R)})|\chi\rangle\langle\chi|(\theta_{A(R)}-P_{A(R)})\\ \label{chip2}
  &\geq (1-\eta)(4\pi a)^2
  \frac{P_{A(R)}}{(2R+1)^3}+\|\chi\|^2_2\left(1-\eta^{-1}\right)(\theta_{A(R)}-P_{A(R)})
  \,.
\end{align}
Here, we have again used the fact that $\sum_x \chi(x) = 4\pi a$.

To conclude the proof, we have to show that $\|\chi\|^2_2\leq \const
a^2/R^3$ for large $R$. This follows from the fact that
$$
\chi(x) \leq \frac 2{R-\widetilde R} \sup_{|x|\geq \widetilde R,
  |e|=1} |\varphi(x) - \varphi(x+e)| \leq C a / R^3
$$
for large $R$, as can be seen from the asymptotic behavior
(\ref{decpp}).  Inserting the inequalities (\ref{chip2}) and (\ref{xxc}) into
(\ref{inequB}), we arrive at the desired result (\ref{tem2}).
\end{proof}

If $|y_1-y_2|> 2\sqrt 3 R$, the cubes of side length $2R$ centered
at $y_1$ and $y_2$, respectively, are disjoint. Hence we can obtain
the following corollary of Lemma~\ref{lemmamain}.

\begin{cor}\label{thecor}
  Let $U$, $W$ and $V$ be as in Lemma~\ref{lemmamain}, and let $U_y =
  T_y U T_y^\dagger$, $W_y= T_yU T_y^\dagger$ and $V_y= T_y V
  T_y^\dagger$, where $T_y$ is the translation operator $(T_y \psi)(x) =
  \psi(x-y)$.  If $y_1,\dots y_n$ satisfy $|y_i-y_j|>2\sqrt{3}R$
  for all $i\neq j$, then
\begin{equation}\label{maincor}
- \Cod{h} \Delta \Co{h} + \frac g2 \sum_{i=1}^n \delta_{x,y_i}
 \geq 4\pi a\sum_{i=1}^n\left[(1-\eps)(1-\eta)U_{y_i} -\frac{W_{y_i}}{\eps}- \frac{CV_{y_i}}{\eta}\right] \,.
\end{equation}
\end{cor}

\subsection {Lemma Two}\label{sec:32}

Recall that the Hilbert space $\mathcal H(N_u,N_d)$ is the subspace of
$L^2(\Z^{3(N_u+N_d)})$ of functions that are separately antisymmetric
in the $N_u$ spin-up variables and the $N_d$ spin-down variables.  For
$\Phi\in\mathcal H(N_u,N_d)$, let $\gamma_u$ and $\gamma_d$ denote the
reduced one-particle density matrices of $\Phi$ for the spin-up and
spin-down particles, respectively, with $\Tr\gamma_u = N_u$ and
$\Tr\gamma_d=N_d$.

For $m\in \Z^3$, we define the functions $f_{m}(x) \in L^2(\Z^3)$ as 
\begin{equation}\label{eigende}
f_{m}(x) = (L+1)^{-3/2}\exp \left (2\pi i m\cdot x (L+1)^{-1}\right) \theta_{[0,L]^3}(x)\,.
\end{equation}
Note that the $f_m$'s are orthonormal functions. For any function $\psi$ supported on $[1,L-1]^3$,
the expectation value $\langle \psi|-\Delta|\psi\rangle$ can be expressed as
\begin{equation}\label{fsp}
  \langle \psi|-\Delta|\psi\rangle=\sum_{m\in[-L/2,(L+1)/2]^3} E\left(2\pi m/(L+1)\right) \left|\langle \psi|f_{m}\rangle\right|^2\,.
\end{equation}

For $M\in \N$, let $\xi(M)$ denote the  projection
\begin{equation}\label{defpm}
\xi(M)=\sum_{ E(2\pi m/(L+1))\leq E_{\rm f}(M/(L+1)^3)} |f_{m}\rangle\langle f_{m}|\,.
\end{equation}
It is easy to see that 
\begin{equation}\label{p1}
\lim_{M\to \infty,L\to \infty} M^{-1}\Tr\, \xi(M)=1
\end{equation}
in the thermodynamic limit $L\to \infty$, $M\to \infty$ with $M/L^3\to \rho$ for some $0< \rho \leq 1$. 

Let $\xi(N_u) = \xi_u$ and $\xi(N_d)=\xi_d$ for simplicity.  As
mentioned in the Introduction, we shall show in Lemma~\ref{lemP} that
the reduced one particle density matrix $\gamma_{u(d)}$ of a ground
state $\Phi$ of $H$ is close to $\xi_{u(d)}$ in an appropriate sense.

\begin{lem}\label{lemP}
Let $\Phi\in \mathcal H_{N_u,N_d}$. Assume that, in the thermodynamic limit $N_{u(d)}\to\infty$,
$L\to\infty$ with $\rho_{u(d)}=N_{u(d)}/L^3$ fixed, 
\begin{equation}\label{assu}
  \limsup_{L\to\infty}\frac 1{L^3} \left\langle \Phi\left| -\mbox{$\sum_{i=1}^{N_u}$}\Delta_{x_i} - \mbox{$\sum_{i=1}^{N_d}$}\Delta_{y_i}\right|\Phi\right\rangle\leq 
  e_0(\rho_u,\rho_d)+ C  a \rho^2
\end{equation}
for some $C>0$ independent of $a$ and $\rho$. Then
\begin{equation}\label{cond}
\limsup_{L\to\infty} \frac 1 {L^3} \Tr[ \gamma_{u(d)}(1-\xi_{u(d)})] \leq \const 
\rho \sqrt { a \rho^{1/3} }. 
\end{equation}
\end{lem}

\begin{proof} The proof is parallel to the proof of Lemma~5 in \cite{LSS}, following an argument in 
\cite{GS}.
\end{proof}

\subsection {Lemma Three}\label{sec:33}

For given points $y_1\ldots y_{N_d} \in \Z^3$, let $I_R(y_1, . . . ,
y_{N_d})$ denote the number of $y_i$'s whose distance to the nearest
neighbor is less or equal to $2\sqrt 3 R$. Because
Corollary~\ref{thecor} can be applied only to those $y_i$'s that
stay away a distance larger than $2\sqrt 3 R$ from all the other
particles, we will need an upper bound on the expectation value of
$I_R$ in the ground state of the Hubbard Hamiltonian. This will be
accomplished in Lemma~\ref{lemir} below. It states that as long as
$R$ is much less than the average particle distance $\rho^{-1/3}$, the
expectation value of $I_R$ in the ground state of $H$ is
small compared to the total number of particles $N$.

\begin{lem}\label{lemir}
Let $\Phi\in \mathcal H(N_u,N_d)$. Assume that for some constant
  $C>0$, independent of $\rho_u$ and $\rho_d$, 
\begin{equation}\label{as3} 
\frac 1{N} \left\langle
    \Phi\left| - \mbox{$\sum_{i=1}^{N_d}$} \Delta_i \right|
    \Phi\right\rangle \leq C \rho^{2/3}\,.  
\end{equation} 
Then 
\begin{equation}\label{ineqI} \langle
  \Phi| I_R(y_1,\dots,y_{N_d}) | \Phi\rangle \leq \const N
  ((R+1)^3\rho)^{2/5}\,.
\end{equation}
\end{lem}

\begin{proof}
  With $D_i$ denoting the distance of $y_i$ to the nearest neighbor
  among the $y_k$ with $k\neq i$, we can write
$$
I_R(y_1,\dots,y_{N_d}) =   \sum_{i=1}^{N_d} \theta_{[0,2\sqrt3R]}(D_i)\,.
$$
Here, $\theta_{[0,2\sqrt3R]}$ denotes the characteristic function of
$[0,2\sqrt3R]$.  It follows from Lemma~\ref{lemlt} below that
\begin{equation}\label{remb}
  \sum_{i=1}^{N_d} \theta_{[0,2\sqrt3R]}(D_i) \leq -b \sum_{i=1}^{N_d} \Delta_{y_i} + b^{-3/2} \frac{2^{11/2}}{15\pi^2} N_d  \sum_{x\in \Z^3} \theta_{[0,2\sqrt3R]}(|x|)
\end{equation}
for any $b>0$.  Using the assumption (\ref{as3}) and optimizing over
the choice of $b$, we arrive at the result.
\end{proof}

It remains to prove the bound (\ref{remb}), which is a special case of
the following Lemma.  Its proof uses a similar decomposition method as in \cite{LT}.

\begin{lem}\label{lemlt}
  Let $f$ be nonnegative, with $\sum_{x\in\Z^3} f(|x|)< \infty$, and
  let $D_i$ denote the distance of $x_i$ to the nearest neighbor among
  all points $x_j$ with $j\neq i$. On the subspace of antisymmetric
  $N$-particle wavefunctions in $L^2(\Z^{3N})$,
\begin{equation}
\sum_{i=1}^N \left( -\Delta_i - f(D_i)\right) \geq - \frac {2^{11/2}}{15 \pi^2} N   \sum_{x\in \Z^3} f(|x|)^{5/2}\,.
\end{equation}
\end{lem}

\begin{proof}
Let $N_1$ be the largest integer less or equal to $N/2$, and let $N_2
= N-N_1$. Consider a partition $P = (\pi_1,\pi_2)$ of the integers
${1, ...,N}$ into two disjoint subsets with $N_1$ integers in $\pi_1$
and $N_2$ integers in $\pi_2$. For a given $P$ and $i\in \pi_1$, we define
$$
D_i^P =  \min\{|x_i-x_j|, j\in\pi_2\}\,.
$$
It is easy to see that 
$$
\sum_{i=1}^N f(D_i) \leq 4 \binom{N}{N_1}^{-1} \sum_P \sum_{i\in\pi_1} f(D_i^P)\,,
$$
where the sum runs over all $\binom{N}{N_1}$ partitions of
$\{1,\dots,N\}$. This follows from the fact that for given $i$ and
$j$, the probability that a partition $P$ has the property that $i\in
\pi_1$ and $j\in \pi_2$ equals $N_1 N_2 / (N(N-1)) > 1/4$.

In particular, we see that 
$$
\sum_{i=1}^N \left( -\Delta_1 - f(D_i)\right) \geq \binom{N}{N_1}^{-1}
\sum_P \sum_{i\in \pi_1} \left(-\frac{N}{N_1} \Delta_i - 4
  f(D_i^P)\right)\,.
$$
For fixed $x_j$, $j\in \pi_2$, we can use the Lieb-Thirring estimate \cite[Thm.~5.3]{DB}
to conclude that 
$$
\sum_{i\in \pi_1} \left(-\frac{N}{N_1} \Delta_i - 4 f(D_i^P)\right)
\geq - \frac 8{15 \pi^2} \left( \frac {N_1}N\right)^{3/2} 4^{5/2}
\sum_{x_1\in \Z^3} f(D_1^P)^{5/2}\,.
$$
Note that the antisymmetry of the wavefunctions is essential here. We can estimate 
$$
\sum_{x_1\in \Z^3}  f(D_1^P)^{5/2} \leq N_2 \sum_{x\in \Z^3} f(|x|)^{5/2}\,.
$$
Using in addition that $N_1^{3/2} N_2 \leq N/ 2^{5/2}$, we arrive at the statement.
\end{proof}

\section{Proof of Theorem~\ref{thm1}}\label{sec:proof}

We write the Hamiltonian $H$ in (\ref{defH}) as 
\begin{equation}\label{hamxy}
  H =\left(-\sum_{i=1}^{N_u}\Delta_{x_i}+\half g \sum_{i=1}^{N_u}\sum_{j=1}^{N_d}
    \delta_{x_i,\y_j}\right) +
  \left(-\sum_{i=1}^{N_d}-\Delta_{y_i}+\half g\sum_{i=1}^{N_u}\sum_{j=1}^{N_d}
    \delta_{x_i,y_j}\right)\,.  
\end{equation} 
Recall that we restrict $H$ to functions that are antisymmetric both
in the $x$ and the $y$ variables, and are supported on the cube
$[1,L-1]^{3(N_u+N_d)}$.  In the following, we are going to derive a
lower bound only on the first term. The lower bound on the second term
can be obtained in the same way by simply exchanging the role of $x$
and $y$. Our bound is not a bound on the ground state energy of this
first term, however, but rather estimates the expectation value of
this term in the ground state of the full Hamiltonian $H$.

First, as mentioned in the Introduction, we decompose the kinetic
energy $-\Delta$ into a high and a low momentum part. Let again
$E_{\rm f}(\rho)$ denote the Fermi energy of an ideal gas of spinless
fermions at density $\rho$, defined in (\ref{deffe}), and let
$$
\widehat M(p)= \sqrt{ \left[ 1 - \frac {E_{\rm f}(\rho_u)}{E(p)} \right]_+ }\,.
$$
Here, $[\,\cdot\,]_+ = \max\{\,\cdot\,, 0\}$ denotes the positive
part.  Moreover, let $\widehat M'(p)= \sqrt{1-\widehat M(p)^2}$. As
pointed out in Eq.~(\ref{momsep}), we can decompose the Laplacian as
$$
\Delta = \Cod{M'}\Delta \Co{M'} + \Cod{M} \Delta \Co{M}\,.
$$

We first claim that  
\begin{equation}\label{lyclaim}
  \lim_{L\to\infty}\frac{1}{L^3} \infspec 
\left[ -\sum_{i=1}^{N_u} \left( \Cod{M'} \Delta \Co{M'}\right)_i \right] \geq e(\rho_u) \,.
\end{equation}
The proof follows in exactly the same way as the proof of Eq.~(64) in
\cite{LSS}, using an argument in \cite{liyau}.

We proceed with the high-momentum part. 
Let $l : \R^3 \to \R_+$  be a smooth, radial, positive
function, with $l(p)=0$ for $|p|\leq 1$, $l(p)=1$ for $|p|\geq 2$, and
$0\leq l(p)\leq 1$ in-between. As in \cite{LSS}, we choose $\widehat h_s(p)$ as
\begin{equation}\label{defchi}
\widehat h_s(p)=l(s p)\,.
\end{equation}
Since $h_s(p)= 0$ for $|p|\leq 1/s$, we can estimate
\begin{equation}\label{Mh}
  \widehat M(p)^2 = \left[ 1 - \frac{E_{\rm f}(\rho_u)}{E(p)}\right]_+  
\geq \left[ 1- \frac{E_{\rm f}(\rho_u)}{\min_{|p|\geq 1/s} E(p)}\right]_+ \widehat h_s(p)^2\, .
\end{equation}
(Here, the minimum is taken over $p\in [-\pi,\pi]^3$, $|p|\geq 1/s$.)
In particular, this implies that
\begin{equation}\label{implcor}
-\Cod{M} \Delta \Co{M} \geq  \left[ 1- \frac{E_{\rm f}(\rho_u)}{\min_{|p|\geq 1/s} E(p)}\right]_+ \left( -\Cod{h_s} \Delta \Co{h_s}\right)\,.
\end{equation}
Since $E(p) \sim |p|^2$ for small $|p|$, and $E_{\rm f}(\rho_u) \leq \const \rho_u^{2/3}$, we obtain
\begin{equation}\label{ses}
 \left[ 1- \frac{E_{\rm f}(\rho_u)}{\min_{|p|\geq 1/s} E(p)}\right]_+ \geq 1 - \const s^2 \rho_u^{2/3}
\end{equation}
as long as $s \gg 1$. 

We note that with this choice of $\widehat h_s(p)$ the corresponding
$\widehat h_s'(p)=1-\widehat{h_s}(p)$ is a smooth function that is 
supported in $|p|\leq 2s^{-1}$. As in \cite{LSS}, we conclude that the
corresponding potential $W(x)$ defined in Lemma~\ref{lemmamain}
satisfies (for $1 \leq R\leq \const s$)
\begin{equation}\label{intw} 
W(x) \leq \const \frac{R^2}{s^5} \ \quad {\rm and\quad } \sum_{x\in \Z^3} W(x) \leq
\const \frac {R^2}{s^2} 
\end{equation}
for some constants depending only on the choice of $l$. Moreover, if
$|y_i-y_j|> 2\sqrt 3 R$ for all $i\neq j$, then
\begin{equation}\label{normw} 
\sum_{i=1}^{N_d} W_R(x-y_i) \leq \const \frac 1{R  s^2} 
\end{equation}
independent of $x$ and $N_d$.

We will now use Corollary~\ref{thecor} to get a lower bound on the sum
of the high momentum part of the kinetic energy and the interaction
energy.  In order to be able to apply this corollary, we have to
neglect the interaction of the $x$ particles with those $y$ particles
that are \textit{not} at least a distance $2\sqrt 3 R$ from the other
$y$ particles.  Let us denote $Y = (y_1,\dots,y_{N_d})$, and let $Y'$
be the subset of $Y$ containing those $y_i$ whose distance to all the
other $y_j$'s is larger then $2\sqrt 3 R$.  Note that, by definition,
the cardinality of $Y'$ is $|Y'|= N_d-I_R(y_1,\dots,y_{N_d})$, with
$I_R$ defined in Section~\ref{sec:33}. Moreover, let $Y''\subset Y'$
be the set of $y_j\in Y'$ whose distance to the boundary of $[0,L]^3$
is at least $R+1$. As argued in \cite{LSS}, $|Y''| \geq |Y'| - \const
(L/R)^2$ and hence, in particular,
\begin{equation}\label{YYp} 
  \lim_{L\to
    \infty}L^{-3}\langle\Phi\left|(|Y'|-|Y''|)\right|\Phi\rangle=0 
\end{equation}
in the ground state $\Phi$ of $H$.

Applying Corollary~\ref{thecor}, together with (\ref{implcor}), we obtain that for a given configuration of $Y$, 
\begin{equation}\label{highb}
  - \sum_{i=1}^{N_u} \left[  \Cod{M} \Delta \Co{M} \right]_i  + \frac g2 \sum_{i,j} \delta_{x_i,y_j} \geq 
  \left[ 1- \frac{E_{\rm f}(\rho_u)}{\min_{|p|\geq 1/s} E(p)}\right]_+ \sum_{i=1}^{N_u} \left[ w_{Y} \right]_i \,,
\end{equation}
with $w_{Y}$ defined as
\begin{equation}\label{defwy}
  w_Y = \sum_{\{ j\, : \, y_j \in Y''\}} 4\pi a \left( (1-\eps)(1-\eta) U_{y_j} - \frac {W_{y_j}}\eps -\frac{CV_{y_j}}{\eta}\right) \,.
\end{equation}

For any $\Phi\in\mathcal H(N_u,N_d)$, we can express the expectation value of $\sum_i [w_Y]_i$ as 
\begin{equation}\label{wphi}
  \left\langle \Phi\left| \sum_{i=1}^{N_u} \left[w_Y\right]_i \right| \Phi\right\rangle = \sum_Y \rho_Y\, \Tr[\gamma_Y w_Y] \,,
\end{equation}
where we denote by $\rho_Y$ the distribution function of $Y=(y_1\cdots y_{N_d})$, that is, 
$$
\rho_Y = \langle\Phi |\delta_{Y=\{y_1\ldots y_{N_d}\}}|\Phi \rangle
$$
and $\gamma_Y$ denotes the one-particle density matrix of $\Phi$ for {\it fixed} $Y$, i.e.,
$$
\gamma_Y(x,x') = \frac {N_u}{\rho_Y} \sum_{x_2,\dots,x_{N_u}}
\Phi(x,x_2,\dots,x_{N_u},Y) \Phi(x',x_2,\dots,x_{N_u},Y)^*\,.
$$
Note that $0\leq \gamma_Y\leq 1$ and $\Tr \gamma_Y = N_u$. Moreover,
$\sum_Y\rho_Y = 1$ and $\sum_Y \rho_Y \gamma_Y = \gamma_u$, which is
the one-particle density matrix for the spin-up particles introduced earlier in Section~\ref{sec:32}.

From now on, we will consider $\Phi$ to be a ground state of $H$. The
assumptions of Lemma~\ref{lemP} and~\ref{lemir} are clearly satisfied for this
$\Phi$, as the upper bound derived in \cite{A} shows. 

Recall that $\xi_u$ denotes the projection $\xi(N_u)$ defined in
(\ref{defpm}).  We write $w_Y =w_{Y,+}-w_{Y,-}$, where $w_{Y,+} \geq
0$ stands for the part of $w_Y$ in (\ref{defwy}) containing $U$,
whereas $w_{Y,-}\geq 0$ is the part of $w_Y$ containing $W$ and
$V$. As proved in \cite[Sect.~V.C]{LSS} we have, for any $\delta>0$,
\begin{align}\label{YY} 
\Tr[\gamma_Y w_Y] &\geq \Tr[\xi_uw_{Y,+}](1-\delta) - \Tr[\xi_uw_{Y,-}](1+\delta) \\\nonumber
& \quad  - \left(1+\delta^{-1}\right) \left( \|w_{Y,+}\|+\|w_{Y,-}\|\right)
\Tr[\gamma_Y(1-\xi_u)] - \|w_Y\| \Tr[\xi_u(1-\gamma_Y)]\,. 
\end{align}

We are now going to bound the various terms on the right side of
(\ref{YY}). First, using $\sum_{x\in \Z^3} U(x) = 1$ and the fact that
$\xi_u$ has a constant density, we have
$$
\Tr[\xi_uw_{Y,+}]  = \frac {\Tr[\xi_u]}{(L+1)^3}  (1-\eps)(1-\eta)4\pi a  |Y''| \,.
$$
To estimate $\lim_{L\to \infty}L^{-3} \sum_Y \rho_Y\Tr[\xi_uw_{Y,+}]$,
we can use (\ref{YYp}), $|Y'|= N_d - I_R(Y)$ and Lemma~\ref{lemir} to
conclude that 
$$
\lim_{L\to\infty}L^{-3} \sum_Y \rho_Y\Tr[\xi_uw_{Y,+}] \geq
(1-\eps)(1-\eta)4\pi a\rho_u\rho_d-\const  a\rho^2((R+1)^3\rho)^{2/5}\,.
$$
Here, we have also used that 
$\lim_{L\to \infty}L^{-3}\Tr[\xi_u]=\rho_u$, which follows from (\ref{p1}).

Analogously, using (\ref{intw}) and the fact that for any $\psi$ and fixed $y$, 
$$
\langle \psi|V_y|\psi\rangle = \frac 1{2(2R+1)^6} \sum_{(x,x')\in
  y+A(R)} |\psi(x)-\psi(x')|^2 \leq \frac 12 \max_{(x,x')\in
  y+A(R)}|\psi(x)-\psi(x')|^2\,,
$$
it is easy to get the upper bound
$$
\lim_{L\to\infty}L^{-3} \sum_Y \rho_Y\Tr[\xi_u w_{Y,-}] \leq \const
 a \rho_u\rho_d \left(\frac{ R^2}{\eps s^2}+\frac{ R^2 \rho^{2/3} }{\eta}\right)\,.
$$
Moreover, using (\ref{normw}) and the fact that the distance between
two $y_j\in Y''$ is at least $2\sqrt 3R$, as well as $\eta \leq 1$, we find that, 
\begin{equation}\label{normw2}
  \|w_Y\| \leq \|w_{Y,+}\|+\|w_{Y,-}\|\leq  \const  a\left( \frac {1}{\eps s^2 R} +\frac {1}{\eta  R^3}\right)\,.
\end{equation}

The bound in Lemma~\ref{lemP} implies that 
$$
\lim_{L\to\infty}L^{-3}\sum_Y \rho_Y \Tr[\gamma_Y (1-\xi_u)] =
\lim_{L\to\infty}L^{-3} \Tr[\gamma_u (1-\xi_u)] \leq \const \rho
(a^3\rho)^{1/6}\,.
$$
Finally, the last term in (\ref{YY}) can be bounded as
$$
\lim_{L\to\infty}L^{-3}\sum_Y\rho_Y\|w_Y\| \Tr[\xi_u(1-\gamma_Y)] \leq
\const a \left( \frac {1}{\eps s^2 R} +\frac {1}{\eta R^3}\right)\rho
(a^3\rho)^{1/6}\,,
$$
where we have used (\ref{normw2}) as well as the fact that
$\Tr[\xi_u(1-\gamma_Y)]=\Tr[\gamma_Y(1-\xi_u)] +
\Tr[\xi_u-\gamma_Y]$. The last term, when averaged over $Y$, is $o(N)$
in the thermodynamic limit, i.e., $\sum_Y \rho_Y \Tr[\xi_u-\gamma_Y] =
o(N)$.

Collecting all the bounds, and applying the same arguments also
to the second term in (\ref{hamxy}), we arrive at the lower bound
\begin{align}\nonumber
  & \lim_{L\to\infty} \frac 1{L^3} E_0(N_u,N_d,L) \\ \nonumber & \geq
  e_0(\rho_u,\rho_d)+ 8\pi a \rho_u \rho_d \left[ 1 -\eps -\eta-\delta
    - \const \left( s^2 \rho^{2/3}+ \frac{R^2}{\eps s^2}
      +\frac{R^2\rho^{2/3}}{\eta }\right) \right. \\ \nonumber &
  \qquad\qquad\qquad\qquad\qquad \left. - \const \left(
      (R^3\rho)^{2/5} +\frac{ (a^3\rho)^{1/6}}\delta \left( \frac
        {1}{\eta R^3\rho} + \frac {1}{\eps s^2 R\rho}\right) \right)
  \right]\,.
\end{align}
Here, we have assumed that $R\geq 1$ and that  $s\gg 1$ in order to be able to apply the
estimate (\ref{ses}).  If we choose
\begin{equation}\label{chos}
R=\rho^{-1/3} (a^3\rho)^{1/30} \ , \ s=\rho^{-1/3}(a^3\rho)^{1/90}\ , \  \eps=\delta= \eta=(a^3\rho)^{1/45}
\end{equation}
this implies that 
$$
\lim_{L\to\infty} \frac 1{L^3} E_0(N_u,N_d,L) \geq e_0(\rho_u,\rho_d)
+ 8\pi a \rho_u\rho_d\left( 1 - \const \big(a\rho^{1/3}\big)^{1/15}\right)\,,
$$
which is the result stated in (\ref{mainresult}).

Recall that in order to be able to apply Lemma~\ref{lemmamain}, it is
necessary that $R\geq C$ for some constant $C>0$, which for our choice of $R$ in (\ref{chos})
is the case if
\begin{equation}\label{sa}
\big(a\rho^{1/3}\big)^{1/10} \geq C \rho^{1/3}\,.
\end{equation}
Under this assumption, also $s\gg 1$ is satisfied for small $a^3\rho$.
For fixed $a$, (\ref{sa}) holds for small enough $\rho$. However, if
$a$ is very small, (\ref{sa}) is violated. In this case, one can
obtain our main result (\ref{mainresult}) actually much easier. One
simply omits the use of Lemma~\ref{lemmamain} altogether, and
applies our perturbative estimate (\ref{YY}) directly to the
interaction potential $(g/2) \delta_{xy}$. Notice that for small $a$,
$g\sim 8\pi a$ is also small. The resulting bound is
\begin{equation}
 \lim_{L\to\infty} \frac 1{L^3} E_0(N_u,N_d,L) \geq   
 e_0(\rho_u,\rho_d)+  g \rho_u \rho_d   \left[ 1 -\delta  - \const \frac {(a^3\rho)^{1/6}}{\delta \rho}\right]\,.
\end{equation}
Choosing $\delta = \rho^{-1/2} (a^3\rho)^{1/12}$ and noting that $g
\geq 8\pi a$ and $\rho \geq \const (a^3\rho)^{1/10}$ in the parameter
regime we are interested in here, this yields (\ref{mainresult}) in
case (\ref{sa}) is violated.

This finishes the proof of Theorem~\ref{thm1}.

\bigskip {\it Acknowledgments:} We are grateful to Alessandro
Giuliani for helpful discussions and remarks.  R.S. acknowledges
partial support by U.S. NSF grant PHY-0652356 and by an A.P. Sloan
Fellowship.

\end{document}